\documentclass[aps,prl,reprint,superscriptaddress]{revtex4-1}
\usepackage{epsfig}
\usepackage{graphicx}
\usepackage[caption=false]{subfig}
\usepackage{bm}
\usepackage{color}
\usepackage{float}
\usepackage{mathtools}
\usepackage{amsmath}

\begin{document}



\title{Stripe Antiferromagnetism and Disorder in the Mott Insulator NaFe$_{1-x}$Cu$_{x}$As ($x \lesssim 0.5$)}


\author{Yizhou Xin}
\affiliation{Department of Physics and Astronomy, Northwestern University, Evanston IL 60208, USA}
\author{Ingrid Stolt}
\affiliation{Department of Physics and Astronomy, Northwestern University, Evanston IL 60208, USA}
\author{Yu Song}
\altaffiliation{Present address: Department of Physics, University of California, Berkeley CA 94720, USA }
\affiliation{Department of Physics and Astronomy and Rice Center for Quantum Materials, Rice University, Houston TX 77005, USA}

\author{Pengcheng Dai}
\affiliation{Department of Physics and Astronomy and Rice Center for Quantum Materials, Rice University, Houston TX 77005, USA}
\author{W. P. Halperin}
\affiliation{Department of Physics and Astronomy, Northwestern University, Evanston IL 60208, USA}


\date{\today}
\begin{abstract}
Neutron scattering measurements have demonstrated that the heavily Cu-doped NaFe$_{1-x}$Cu$_{x}$As compound behaves like a Mott insulator exhibiting both real space Fe-Cu stripes, as well as antiferromagnetism below a N\'eel temperature for $x\lesssim 0.5$. We have investigated evolution of structural and magnetic ordering using $^{23}$Na and $^{75}$As NMR for single crystals ($x$ = 0.39 and 0.48), confirming antiferromagnetism in the form of magnetic stripes.  We show that end-chain defects in these stripes are the principal source of magnetic disorder and are responsible for cluster spin-glass transitions in both compounds, in the latter case coexistent with antiferromagnetism. Aided by our numerical simulation of the $^{75}$As spectra, we show that a staggered magnetization at the Fe sites is induced by non-magnetic Cu dopants.

\end{abstract}

\pacs{}

\maketitle

\begin{figure}
	\includegraphics[scale = 0.45]{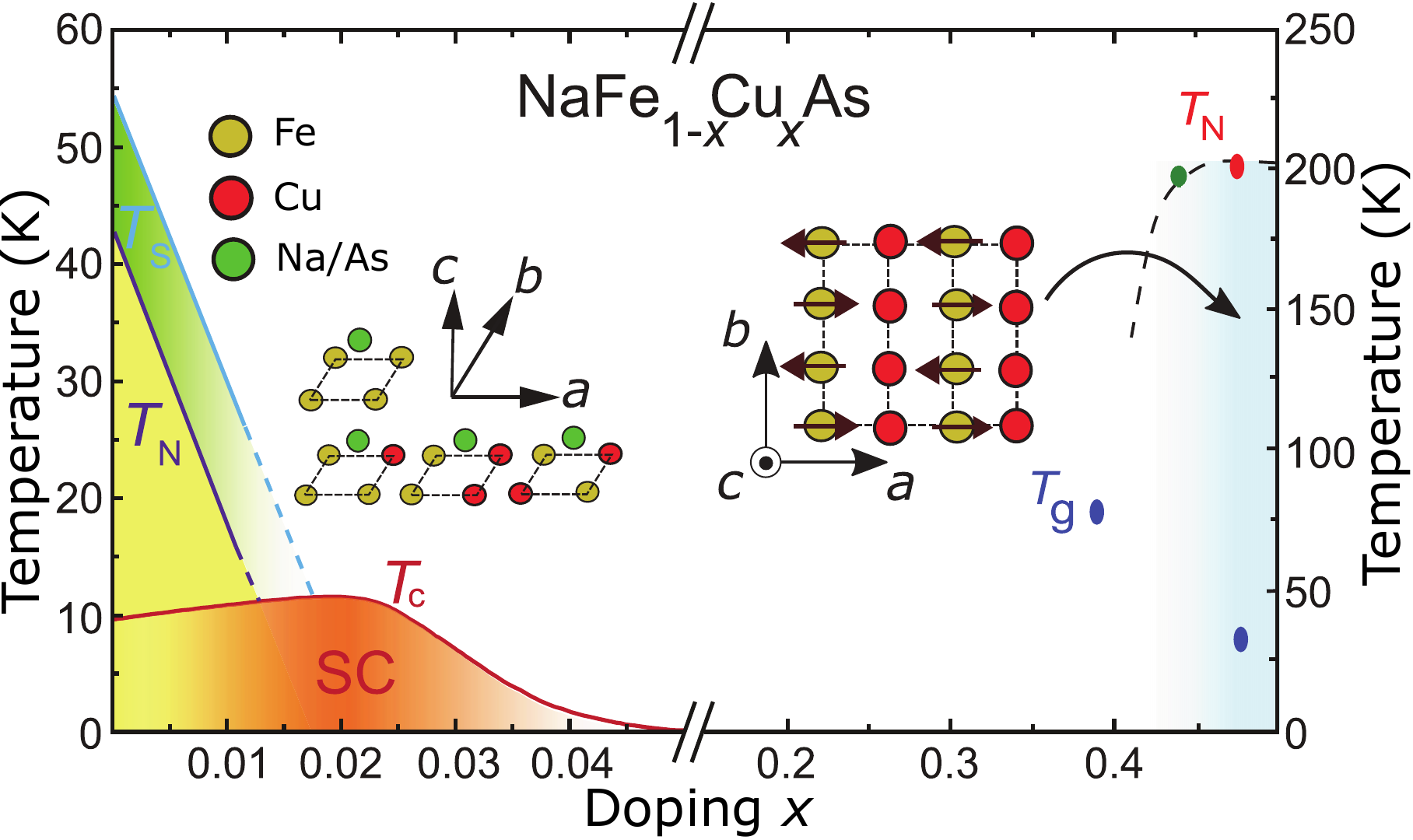}
	\caption{\label{Fig.1} Phase diagram of NaFe$_{1-x}$Cu$_{x}$As single crystals showing superconducting regions at low Cu doping and magnetic ordering at high doping levels; adapted from Ref.~\cite{Wan.13} for $x< 0.05$. The green data point is from neutron scattering Ref.~\cite{Son.15}; red and blue points are from this work. $T_{N}$ is the N$\acute{e}$el temperature and $T_g$ spin-glass transitions from NMR.  Insets show Na and As sites relative to nearest-neighbor Fe or Cu substituted for Fe.}
\end{figure}

An important step in understanding the physics of iron-based superconductors is to investigate the connection between superconductivity and magnetism. The heavily Cu-doped pnictide, NaFe$_{1-x}$Cu$_x$As (phase diagram displayed in Fig.~\ref{Fig.1}) becomes a Mott insulator that exhibits both real space Fe-Cu stripe ordering and  long-range antiferromagnetism (AFM) below the N\'eel temperature $T_{N}\approx200$\,K, for $x$ close to 0.5~\cite{Son.16,Mat.16,Yu.13}. Later work shows the importance of the interplay of electronic correlations and  spin-exchange coupling~\cite{Zha.17,Cha.17}. This is the only known Fe-based material for which superconductivity can be smoothly connected to a Mott-insulating state with increasing doping. Our recent investigation~\cite{Xin.19} has shown a systematic development of AFM Fe-Cu clusters  with increasing Cu dopant at low copper concentrations. In this paper we report nuclear magnetic resonance (NMR) and magnetization measurements complemented by numerical simulation for $x\geq0.39$ that identify antiferromagnetic and cluster spin-glass transitions, and staggered magnetization induced by non-magnetic Cu even in the paramagnetic state. Our results reveal stripe AFM and structural evolution, as well as the link between them, consistent with neutron scattering~\cite{Son.16}. With NMR we have discovered coexistence of long-range AFM and cluster spin-glass that correspond to end-chain defects, at temperatures $T\lesssim30$ K for $x = 0.48$, making the compound the first iron pnictide system that shows coexistence of long-range magnetic order and spin-glass behavior.

As indicated in Fig.~\ref{Fig.1}, the $^{75}$As and $^{23}$Na nuclei in NaFe$_{1-x}$Cu$_{x}$As are located on opposite sides of the Fe layer and are both coupled to the electronic spins at four nearest-neighbor (NN) sites in the Fe-Cu plane via transferred hyperfine interaction. Our $^{23}$Na NMR spectra, nuclear spin-lattice ($1/^{23}\mathrm{T}_1$), and spin-spin ($1/^{23}\mathrm{T}_2$) relaxation rates show evidence of the $x = 0.48$ compound being a stripe-ordered Mott insulator with a N$\mathrm{\acute{e}}$el transition temperature at $T_{N}=200$ K, first identified by neutron scattering~\cite{Son.16,Mat.16,Yu.13} for $x = 0.44$ confirmed from our NMR measurements. 
We find that the long-range three-dimensional AFM order for $x = 0.48$ coexists with a spin-glass phase at lower temperatures, $T\leq30$\,K, due to magnetic frustration in the Fe-Cu plane.  In contrast the compound, $x = 0.39$, forms a cluster spin-glass with a much higher transition temperature, $T_{g}\approx80$ K. Aided by numerical simulation, our analysis of the $^{75}$As NMR spectra for $x\lesssim0.5$, indicates existence of staggered magnetization induced by imperfect stripes of non-magnetic Cu dopants, similar to that in cuprates~\cite{Jul.00,Bob.97,Wal.93, Mor.98}.

Single crystals of NaFe$_{1-x}$Cu$_{x}$As were grown by the self-flux method at Rice University~\cite{Son.16}. Detailed information on sample preparation, including specially designed hermetic sample holders for NMR measurements, can be found elsewhere~\cite{Xin.19}. We performed $^{75}$As and $^{23}$Na NMR experiments for compounds $x$ = 0.39 and 0.48, with the $c$-axis parallel to the external magnetic field, $H_0 = 13.98$ T. We used the quadrupolar echo ($\pi/2$-$\tau$-$\pi/2$) NMR pulse sequence  and a saturation-recovery sequence for 1/$^{23}\mathrm{T}_{1}$, with a $\pi/2$ pulse length $\sim6\,\mu s$. 

\begin{figure}
    \hspace*{-0.5cm}
	\includegraphics[scale = 0.38]{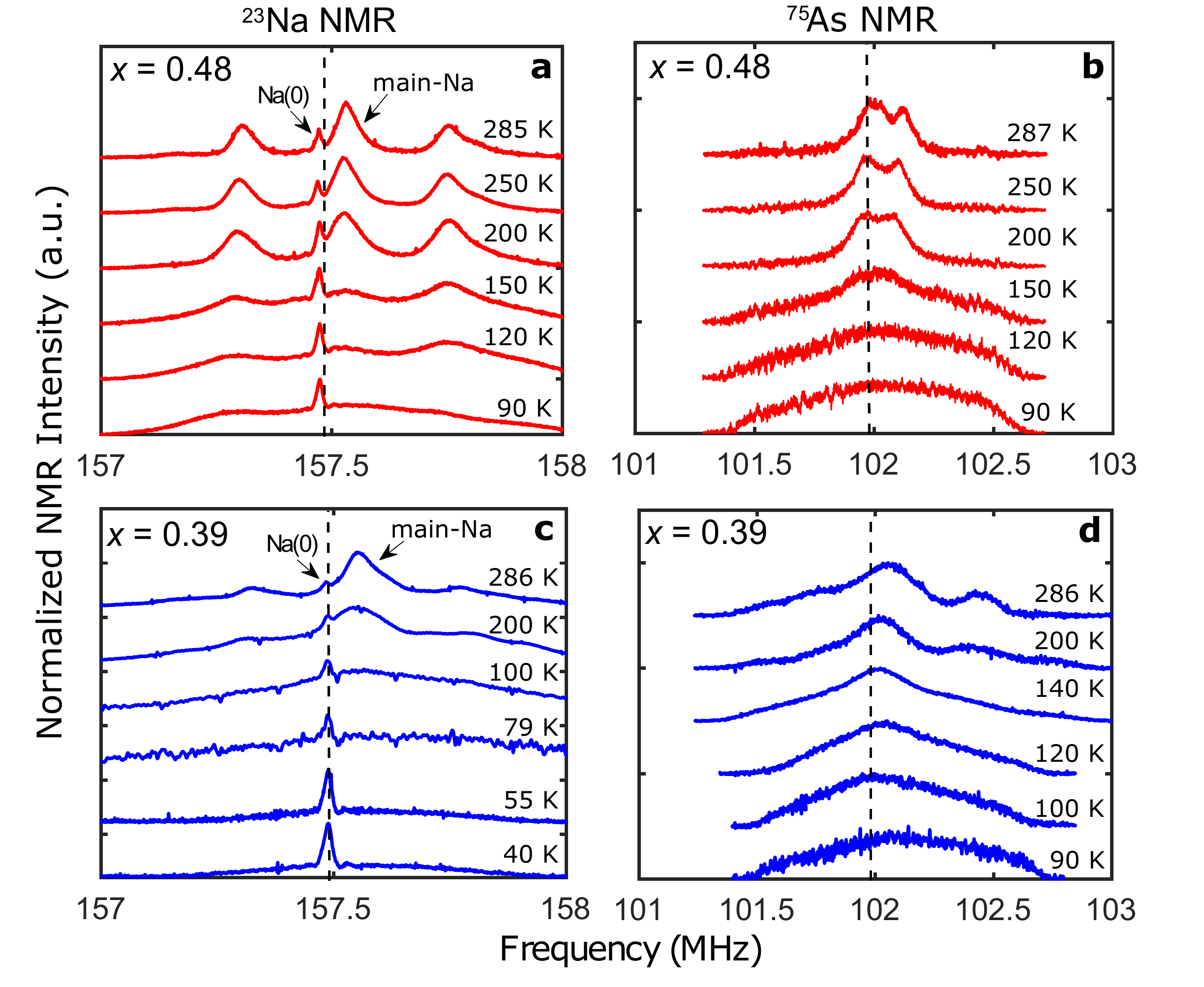}
	\caption{\label{Fig.2} Temperature evolution of $^{23}$Na and $^{75}$As NMR spectra \textbf{\textit{H}}$_{0}$ = 14 T $\parallel$ $c$-axis. (a),(c) $^{23}$Na spectra, left panel; (b),(d) $^{75}$As, right panel; doping $x = 0.48$ is red, 0.39 is blue. There are two inequivalent $^{23}$Na sites, denoted by Na(0) having  no nearest-neighbor Cu sites, otherwise it is main-Na. The room-temperature $^{75}$As spectral splitting, $x = 0.48$ is an end-chain effect (see text). The dashed lines are the Larmor frequencies. All spectra are normalized to peak height.}
\end{figure}

The temperature dependence of the $^{23}$Na spectra for both $x=0.39$ and $0.48$ compositions are shown in Fig.~\ref{Fig.2}\,(a) and (c).  There are two Na spectral components.  One is small and narrow at all temperatures which we have identified with a site that has no NN Cu dopant, denoted by Na(0)~\cite{Xin.19}.  The second, which we call main-Na, appears as a triplet that can be attributed to the central transition and its two quadrupolar satellites for which the satellite splitting is proportional to the local electric-field-gradient (EFG). The main-Na has a significantly larger spectral weight and broader linewidth than Na(0). Its spectral weight and broadening increase with increased doping and with deceasing temperature indicating that it is in a magnetic environment adjacent to one or more Cu atoms substituting for Fe~\cite{Xin.19}. However, for $x = 0.48$ the main-Na peak is narrower, and its quadrupolar satellites are better defined than for $x = 0.39$.  This is consistent with a more uniform structural environment for $x = 0.48$, indicated by a narrower distribution of EFG, and concomitantly, linewidths dominated by the local field distribution.  We ascribe this to the Fe-Cu stripe formation, identified from neutron scattering~\cite{Son.16}. The decrease in total integrated spectral intensity is a `wipe-out' effect, due to both significant inhomogenous broadening of the spectrum, and increased inhomogeneity of the relaxation rates at low temperatures. The latter is discussed in detail in Appendix A.

The temperature dependence of the $^{75}$As spectra for $x=0.48$ and 0.39 is shown in Fig.~\ref{Fig.2}\,(b) and (d). The main $^{75}$As peak that is centered approximately at $\sim102$\,MHz is associated with the As sites that are within the Fe-Cu stripes, evident from its growing relative spectral weight with increasing $x$. A relatively small spectral component centered at $\sim$$102.45$ MHz for $x = 0.39$, is most likely from $^{75}$As sites outside the stripe region and is negligible for $x = 0.48$. The main $^{75}$As peak displays greater inhomogeneous broadening than $^{23}$Na owing to the order-of-magnitude larger hyperfine field coupling to the electronic spins in the Fe-Cu plane~\cite{Xin.19}. Interestingly, for $x \geq 0.39$, the $^{75}$As spectra deviate significantly from that for lower doping (Appendix B) showing  increased magnetic disorder with increasing $x$ at room temperature, but with a marked reversal as $x$ approaches 0.5 which we associate with the stripe order reported from neutron scattering and transmission electron microscopy~\cite{Son.16}.  Remarkably, for the $x=0.48$ composition  the central $^{75}$As peak  shows a splitting of $\sim130$ kHz. The splitting we observe is accurately accounted for by our numerical simulation at various dopings near the fully stripe-ordered compound $x=0.5$ and from these simulations we identify end-chain defects as the source of the splitting.


The temperature dependence of the spin-lattice relaxation rate (1/$^{23}T_{1}$) of the main $^{23}$Na central transition provides important information about spin dynamics. We fit the time dependence of the longitudinal magnetization to a stretched exponential 
 in order to account for a distribution of relaxation rates~\cite{Dio.13,Mit.08,Zon.07,Joh.06}, $M(t) = M_0[1-2f(\theta)(0.9\times e^{-(6t/T_1)^{\beta}}+0.1\times e^{-(t/T_1)^{\beta}})]$, where $f$ is a function of the tipping angle $\theta$.  

\begin{figure}
	\hspace*{-0.5cm}
	\includegraphics[scale = 0.45]{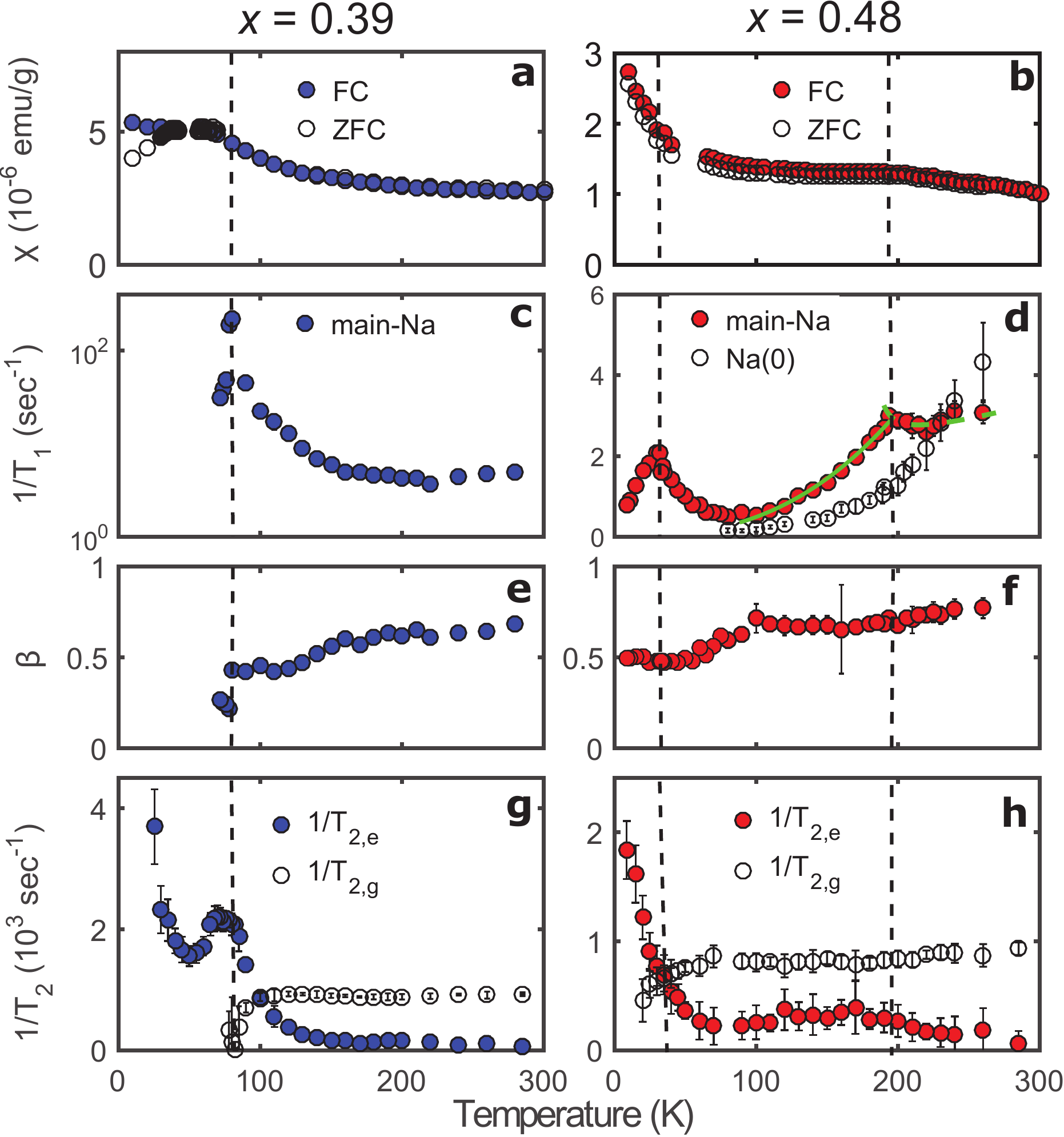}
	\caption{Magnetic transitions (dashed vertical lines). (a),(b) Magnetic susceptibility \textbf{\textit{H}} = 1 T $\parallel$ $ab$-plane. (c),(d) Spin-lattice relaxation rate (1/$^{23}\mathrm{T}_{1}$) of main-Na; curves are fits to data (see text). (e),(f) Stretched exponent $\beta$. (g),(h) Gaussian and exponential components of spin-spin relaxation rate (1/$^{23}\mathrm{T}_{2}$).}\label{Fig.3}
\end{figure}

We find that $x=0.39$ has a glass transition at $T_{g}\approx80$ K, evident from three observations: {\it i}) the bifurcation of magnetic susceptibility $\chi$ vs. $T$ taken under zero-field-cooled (ZFC) and field-cooled (FC) conditions, Fig.~\ref{Fig.3}\,(a); {\it ii}) the peak in 1/$^{23}T_{1}$ vs. $T$ shown in Fig.~\ref{Fig.3}\,(c), and {\it iii}) the peak in 1/$^{23}T_{2,e}$ vs. $T$ in Fig.~\ref{Fig.3}\,(g) with a transition from gaussian to exponential relaxation followed by its increase with lower temperature. Each of these three observations provides an indication of a spin-glass transition.

A stretched exponential form for a relaxation process is a pragmatic approach to account for a distribution in  rates~\cite{Dio.13,Mit.08,Joh.06}. This appears to be the case for both $x=0.39$ and 0.48.  For $x = 0.39$, in the temperature range $T < T_{g}$, $\beta$ has a sudden drop to $\lesssim0.2$, indicating that 1/$^{23}T_{1}$ varies by almost 2 orders of magnitude across the distribution at temperatures $T < T_{g}$~\cite{Joh.06}. In fact, $\beta$ has already started to decrease at $T \approx 150$ K, approaching $\beta \sim0.5$ above the glass transition for both samples, indicating that the inhomogeneity of spin fluctuations is evident well above spin freezing. Magnetic inhomogeneity, described in terms of magnetic cluster formation~\cite{Xin.19}, together with the slow spin dynamics, suggests that $x = 0.39$ is a cluster spin-glass, where local AFM fluctuations develop in disconnected spatial regions with varying domain sizes similar to what has been reported in both underdoped ``122'' pnictide Ba(Fe$_{1-x}$Co$_{x}$)$_2$As$_{2}$ and cuprates~\cite{Dio.13,Cur.00,Jul.99,Mit.08,Bae.12,Wu.13}. 

The temperature dependence of 1/$^{23}\mathrm{T}_{1}$ for Na(0), $x = 0.48$  in Fig.~\ref{Fig.3}\,(d), shows the same pseudogap behavior as with low doping, $x<0.39$, which we have shown to be completely suppressed for the main-Na component~\cite{Xin.19}.  Instead, the temperature dependence of the main-Na rate for $x = 0.48$ has a cusp at $T=200$\,K that can be associated with the N$\mathrm{\acute{e}}$el state transition, in excellent agreement with the results from elastic neutron scattering~\cite{Son.16}. The data are well fit to $T_{1}^{-1} = aT+bT/(T-T_{N})^{1/2}$ shown as a green dashed curve in Fig.~\ref{Fig.3}\,(d), for $T > 200$\,K. The first term represents a Korringa relaxation from itinerant quasiparticles (negligible for $x=0.48$) and the second term arises from 3D fluctuations of AFM local moments~\cite{Mor.74,Dio.10}. The data also shows a power-law behavior of $1/T_{1} \sim T^{2.65\pm0.12}$, close to $T^{3}$, over the intermediate temperature range ($100$\,K $\leq T\leq200$\,K), suggesting a two-magnon Raman process as the main relaxation mechanism that has been observed in an AFM insulating state when $T\gg\triangle$, where $\triangle$ is the  energy gap anisotropy in the spin wave spectrum~\cite{Bee.68,Joh.11}.

At low temperatures near 30\,K there is an additional peak in 1/$^{23}\mathrm{T}_{1}$ for $x = 0.48$  that is evidence for a second transition. It appears that the two magnetic transitions are from distinct regions, one of which forms Fe-Cu stripes and exhibits long-range AFM order at $T_{N}$, while the other is occupied by spin-glass clusters which undergo spin-freezing at $T_{g}$. Supporting this identification, we note that there is similarity in the temperature dependence of $\beta$ and 1/$^{23}\mathrm{T}_{2,g}$  with the glass transition for  $x=0.39$. However, this is in contrast with the temperature independent behavior of the rate at T$_{N}$, ruling out a different form of AFM order at 30\,K. Furthermore, neutron scattering results indicate three-dimensional long-range AFM order for  $T\leq200$\,K, robust down to $\sim4$\,K. We conclude that  there is coexistence of long-range AFM order and cluster spin-glass behavior for $T\leq30$\,K in NaFe$_{0.52}$Cu$_{0.48}$As. Coexistence of spin-glass and long-range order has been discussed theoretically and from numerical simulation~\cite{Rya.92,Kor.87}.

While the temperature dependence of $\chi$ shown in Fig.\ref{Fig.3}\,(b) exhibits no obvious ZFC-FC bifurcation around $T \approx 30$ K, it is possible that it is suppressed by the large external field, H = 1 T.

The temperature dependence of the spin-spin relaxation rate 1/$^{23}{T}_{2}$ is shown in  Fig.~\ref{Fig.3}\,(g) and (h). The exponential ($1/T_{2,e}$) and gaussian ($1/T_{2,g}$) components are extracted from fitting the transverse magnetization, $M(t) = M_0\mathrm{exp}[-\frac{t}{T_{2,e}}]\mathrm{exp}[-\frac{t^2}{T_{2,g}^2}]$. At high temperatures, the relaxation is dominated by 1/$^{23}T_{2,g}$. With $T$ approaching $T_{g}$, both compounds show a crossover from gaussian to exponential decay. The increase of 1/$^{23}T_{2,e}$ at low temperatures is a result of slow spin dynamics due to glass freezing~\cite{Cur.00,Bos.16}. The in-plane resistivity, $\rho_{ab}$, also increases significantly over the same temperature range~\cite{Son.16}, suggesting that charge localization might be a precursor effect to spin freezing, similar to cuprates~\cite{Jul.99}. For $x = 0.39$, the peak in $1/T_{2,e}$ at 80\,K is due in part to a sizable  Redfield contribution from 1/$^{23}T_{1}$. The suppression of $1/T_{2,g}$ indicates quenching of the nuclear spin flip-flop dipole interaction between neighboring $^{23}$Na nuclei where the local field becomes sufficiently inhomogeneous below the glass transition that Zeeman energy is not conserved~\cite{Gen.75,Bos.16}.

\begin{figure}
	\includegraphics[scale = 0.47]{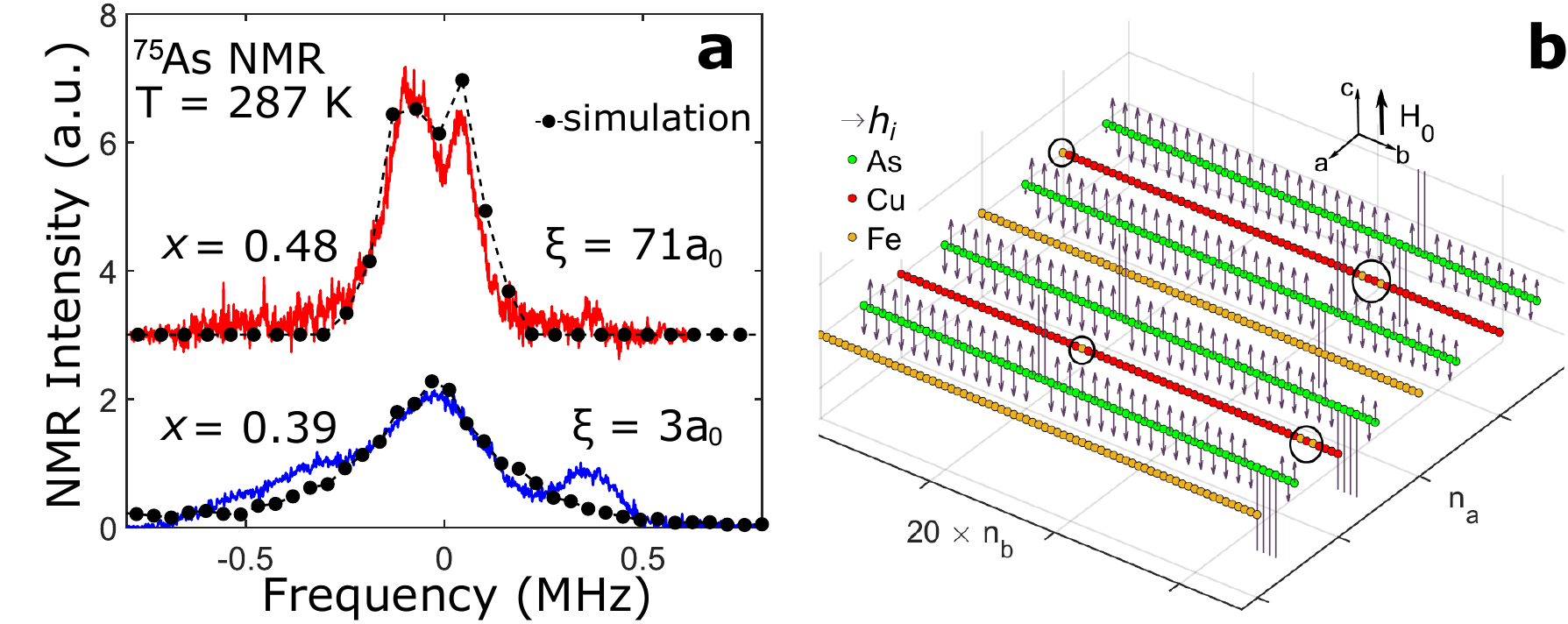}
	\caption{\label{Fig.4} Numerical simulation of room-temperature $^{75}$As spectra with the staggered magnetization model. (a) Comparison between the  $^{75}$As spectra of $x = 0.39$ and 0.48 with simulations. The frequencies of experimental and simulated spectra are aligned for better comparison. (b) Hyperfine fields $\textbf{\textit{h}}$ at the As sites in a simulated Fe-Cu lattice with Fe defects (black circles) in the Cu chains for $x = 0.48$. Only a small portion of the total simulated lattice of the size 400 $\times$ 400 is shown here.}
\end{figure}

A very unusual splitting appears in the $x=0.48$  $^{75}$As spectra for $T\gtrsim T_N$.  At such high temperatures this  distribution in local fields cannot come from static antiferromagnetic  order.  Rather this must be from a distribution of staggered hyperfine fields, \textit{\textbf{$h$}}, at As sites. We describe this as follows.  

The hyperfine field at any As site is a sum of the hyperfine coupling to its NN Fe atoms,

\begin{equation}\label{Eq.1}
\frac{h}{H_0} = A\sum_{i = NN}\frac{{M_i}}{H_0}
= A\sum_{i = NN}{\chi_i^{'}}
\end{equation}

\noindent where $A$ represents the hyperfine coupling between the $^{75}$As nuclear spin and the electronic spins from its NN Fe sites. We have previously measured $A \approx 4.5$ T/$\mathrm{\mu}_B$ for $^{75}$As, with $H||c$~\cite{Xin.19}. The magnetization and local magnetic susceptibility at the NN Fe sites are denoted by $\textit{\textbf{M}}_i$ and $\chi_i^{'}$ respectively. To determine $\chi_i^{'}$ we adopt a gaussian model for the wave vector dependent $\chi^{'}(\textit{\textbf{q}})$~\cite{Jul.00,Bob.97,Wal.93,Mor.98},

\begin{equation}\label{Eq.2}
\chi^{'}(\textit{\textbf{q}}) = 4\pi\chi^{*}(\frac{\xi}{a_0})^2\mathrm{exp}(-(\textit{\textbf{q}}-\textit{\textbf{Q}}_\mathrm{AF})^2\xi^{2})
\end{equation}

\noindent which assumes a peak at \textbf{\textit{Q}}$_\mathrm{AF}$ = (1,1,0) corresponding to the Bragg peak determined from neutron scattering data~\cite{Son.16}. The parameters $\chi^{*}$, $\xi$, and $a_0$ represent the amplitude of the oscillation, AFM correlation length, and lattice constant respectively. The staggered susceptibility is then given by the inverse Fourier transform of Eq.~\ref{Eq.2}~\cite{Bob.97},

\begin{equation}\label{Eq.3}
\chi^{'}(\textbf{r}_i-\textbf{r}_\mathrm{Cu}) = \pm\chi^{*}(-1)^{n_a+n_b}\mathrm{exp}(-|\textbf{r}_i-\textbf{r}_\mathrm{Cu}|^2/4\xi^2)
\end{equation}

\noindent where $\textbf{r}_i -\textbf{r}_\mathrm{Cu} = n_a\mathrm{\textbf{a}}+n_b\mathrm{\textbf{b}}$ represents the relative position of a Fe atom at $\textbf{\textit{r}}_i$ with respect to a Cu dopant, with $n_a$ and $n_b$ specifying the location along the crystalline $a$ and $b$ axes.  
Then to obtain $\chi_i^{'}$ for each Fe, we sum over all Cu sites. The prefactor $(-1)^{n_a+n_b}$ gives rise to the oscillatory behavior of $\chi^{'}$ as a function of position $\textbf{r}_i$ embodying  antiferromagnetic correlations between iron atoms.
We numerically simulated the $^{75}$As lineshape, Fig.~\ref{Fig.4}\,(a), using this staggered magnetization model with a 400 \nolinebreak$\times$ \nolinebreak400 square lattice of Fe-Cu stripes with defects. The defects are Fe atoms on Cu chains introduced at each Cu chain site with a probability $p = 1-2x$, where $x$ is the Cu concentration. Details of the lattice simulation can be found in Appendix D. Thus, the simulated Fe-Cu lattice becomes more stripe-ordered with $x$ approaching $0.5$, while the ratio between Fe and Cu is given by $x/(1-x)$. A comparison between the relative spectral weight of different Na sites given by our simulated lattice and that obtained from $^{23}$Na NMR is shown in the Appendix C.  
We used Eq.\,\ref{Eq.3} to calculate the susceptibility at each Fe site due to all Cu dopants, and then used Eq.\,\ref{Eq.1} 
to compute the hyperfine field along the $c$-axis, \textbf{\textit{h}}, at each As site. The simulated $^{75}$As spectrum is a histogram of these local fields. A least-squares fit of the $^{75}$As spectrum with this simulation gives $\xi = 3a_0$ and $71a_0$, for $x = 0.39$ and 0.48 respectively, leading to the splitting in the simulated spectrum of $x = 0.48$. Shown in Fig.~\ref{Fig.4}\,(b), the alternating \textbf{\textit{h}} at the As sites in the simulated Fe-Cu lattice is responsible for the splitting. Similar simulation results are given for \textit{Q}$_\mathrm{AF}$ = (0,1,$\frac{1}{2}$), as shown in the Appendix E.

Our result of $\xi = 71a_0$ for $x = 0.48$, expected to be roughly temperature independent for $T_{N} \geq 200$ K, is comparable to the weakly temperature-dependent AFM correlation length $\sim97a_0$ from  neutron scatftering of the compound $0.44$ at temperatures below $T_{N}$~\cite{Son.16}. However, it is not established that these two correlation lengths are related.  We emphasize that the staggered magnetization at Fe sites is in the paramagnetic state, fundamentally different from spontaneous long-range AFM order, and can only be revealed by an external field \textbf{\textit{H}}$_{0}$ which was not applied in the neutron scattering measurements. The staggered magnetization we have identified is mediated by valence electrons in an insulating system and rather different from the RKKY susceptibility in a metallic system. A connection can be made with NMR studies of staggered magnetization induced by both non-magnetic and magnetic impurities in cuprates~\cite{Jul.00,Bob.97,Wal.93, Mor.98}, showing that AFM correlations enhanced by non-magnetic dopants are a common phenomenon in both pnicides and cuprates.
  
Since we have found that disorder in Fe-Cu stripes is responsible for the splitting in the simulated $^{75}$As spectrum, we have used the simulation as a tool to investigate other possible structures, where Fe and Cu are either randomly distributed or stripe-ordered without defects. No splitting larger than the simulation resolution $\sim60$\,kHz was found for $\xi\leq200a_{0}$. It is also worthwhile noting that we found no splitting  in our simulated $^{23}$Na spectra, consistent with experiment, which we attribute to the weaker hyperfine coupling of $^{23}$Na: $^{23}\mathrm{A}/^{75}\mathrm{A}\approx$1/12~\cite{Xin.19}.

\begin{figure}
	\includegraphics[scale = 0.30]{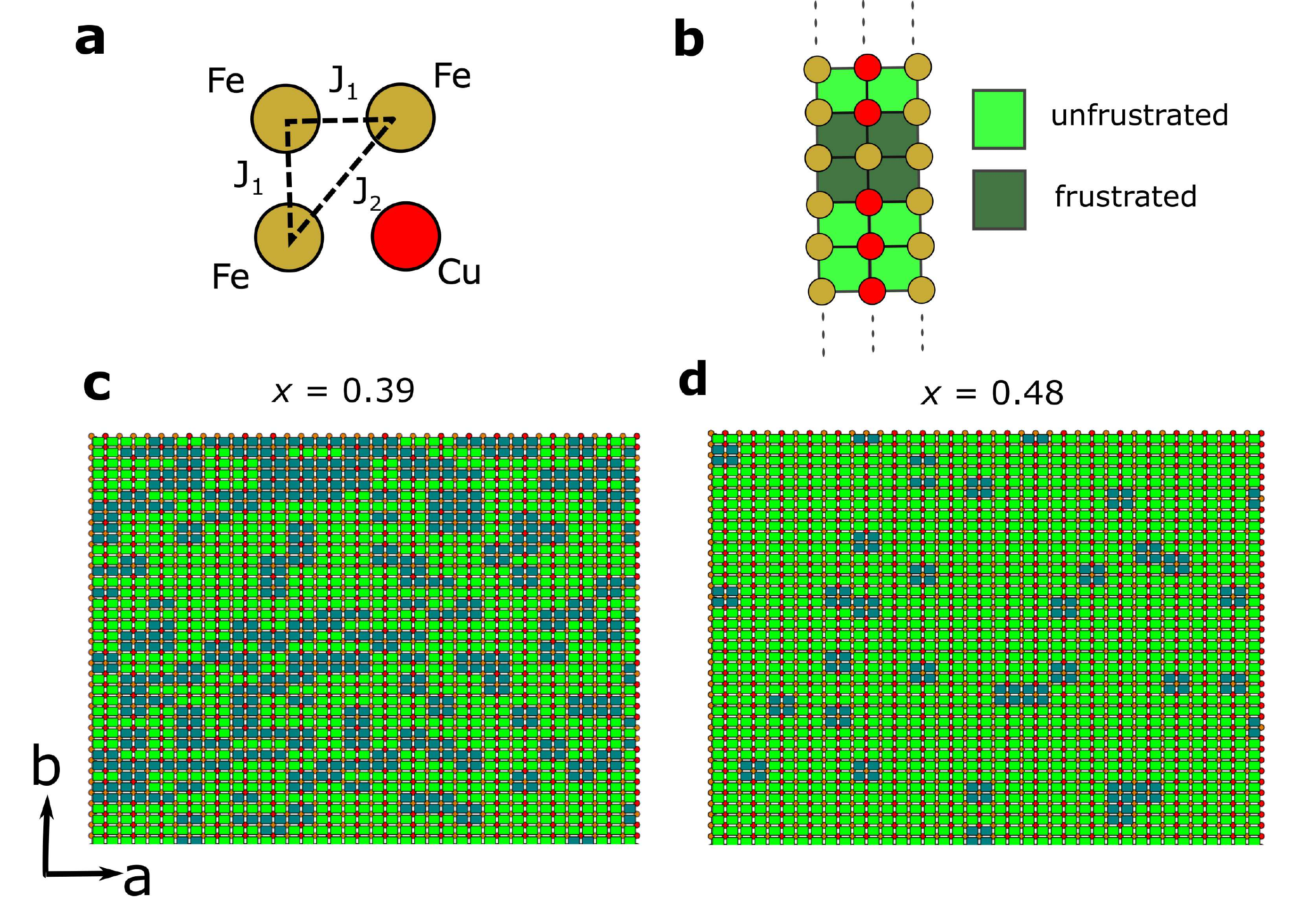}
	\caption{Frustration in the Fe-Cu lattice is due to Fe defects in Cu stripes. (a) Schematic of J$_{1}$ coupling between two nearest-neighbor Fe sites, and J$_{2}$ coupling between two next-nearest-neighbor Fe sites. (b-d) Schematic of a Fe-Cu stripe with Fe defects in the Cu chain (dark green). Frustration is lifted in the region away from defects (light green).  The observed cluster spin-glass transition temperatures are 80\,K for $x=0.39$ and 30\,K  for $x=0.48$.}\label{Fig.5}
\end{figure}

We have demonstrated that insight into the interplay between the structural and magnetic order 
can be gained from visualization of the Fe-Cu lattice using a simulation and we have identified end-chain defects.  It is natural to associate these defects with the magnetic disorder and frustration in the cluster spin-glass states we have observed ~\cite{Edw.75,Zon.07,Bin.86}. This is illustrated in Fig.~\ref{Fig.5}\,(a), where competing exchange interactions between nearest-neighbor Fe atoms ($J_{1}$) and that between next-nearest-neighbor Fe atoms ($J_{2}$) causes magnetic frustration, for which a theoretical basis is discussed in Ref.~\cite{Si.08}. The frustration is lifted by non-magnetic Cu in ideal Fe-Cu stripes~\cite{Wan.17}. However, in the presence of disorder in the stripes, frustration can be locked-in as shown in the schematic in Fig.~\ref{Fig.5}\,(b), where we distinguish a frustrated Fe-Cu square from a non-frustrated one. Compared to $x = 0.39$, the compound $x = 0.48$ shows smaller, but non-negligible frustrated regions, Fig.~\ref{Fig.5}\,(c) and (d), qualitatively consistent with our interpretation of the origin of the low-temperature spin-glass transition being caused by a region occupied by frustrated Fe-Cu clusters coexisting in an antiferromagnetic background.

In summary, with $^{23}$Na NMR we have identified the N$\acute{\mathrm{e}}$el state transition in the compound NaFe$_{0.52}$Cu$_{0.48}$As at $T_{N}=200$ K, and a cluster spin-glass phase at temperatures $T\leq30$\,K owing to disorder in Fe-Cu stripes. For NaFe$_{0.61}$Cu$_{0.39}$As, there is no similar long range antiferromagnetic order; however, we have observed a cluster spin-glass transition at $T \approx 80$\,K in that compound.  Our $^{75}$As NMR spectra for the $x = 0.48$ compound has a well-defined splitting at temperatures above the N$\acute{\mathrm{e}}$el transition temperature.  Aided by numerical simulation we interpret this splitting as evidence for a staggered magnetization induced by non-magnetic Cu in a lattice occupied by Fe-Cu stripes with defects.  

\section{acknowledgment}
We thank Weiyi Wang and Chongde Cao for their contributions to the crystal growth and characterization. The NMR spectrometer, MagRes2000  wide-band spectrometer system, was designed by A. P. Reyes at the National High Magnetic Field Laboratory (NHMFL). The home-built continuous flow cryostat was designed by J.A. Lee at Northwestern University. Research was supported by the U.S. Department of Energy (DOE), Office of Basic Energy Sciences (BES), Division of Material Sciences and Engineering under Award No. DE-FG02-05ER46248 (WPH) and DE-FG02-05ER46202 (PD), and the NHMFL by NSF and the State of Florida. The single crystal growth efforts at Rice were supported by the U.S. DOE, BES under Grant No. DE-SC0012311.  Part of the materials work at Rice was supported by the Robert A. Welch Foundation under Grant No. C-1839.

\newcommand{\beginappendixa}{%
	\setcounter{table}{0}
	\renewcommand{\thetable}{A\arabic{table}}%
	\setcounter{figure}{0}
	\renewcommand{\thefigure}{A\arabic{figure}}%
	\setcounter{equation}{0}
	\renewcommand{\theequation}{A\arabic{equation}}%
}
\beginappendixa
\section{APPENDIX A: `Wipe-out' effect in $^{23}$Na Spectra}

\begin{figure}[h]
	\includegraphics[scale = 0.29]{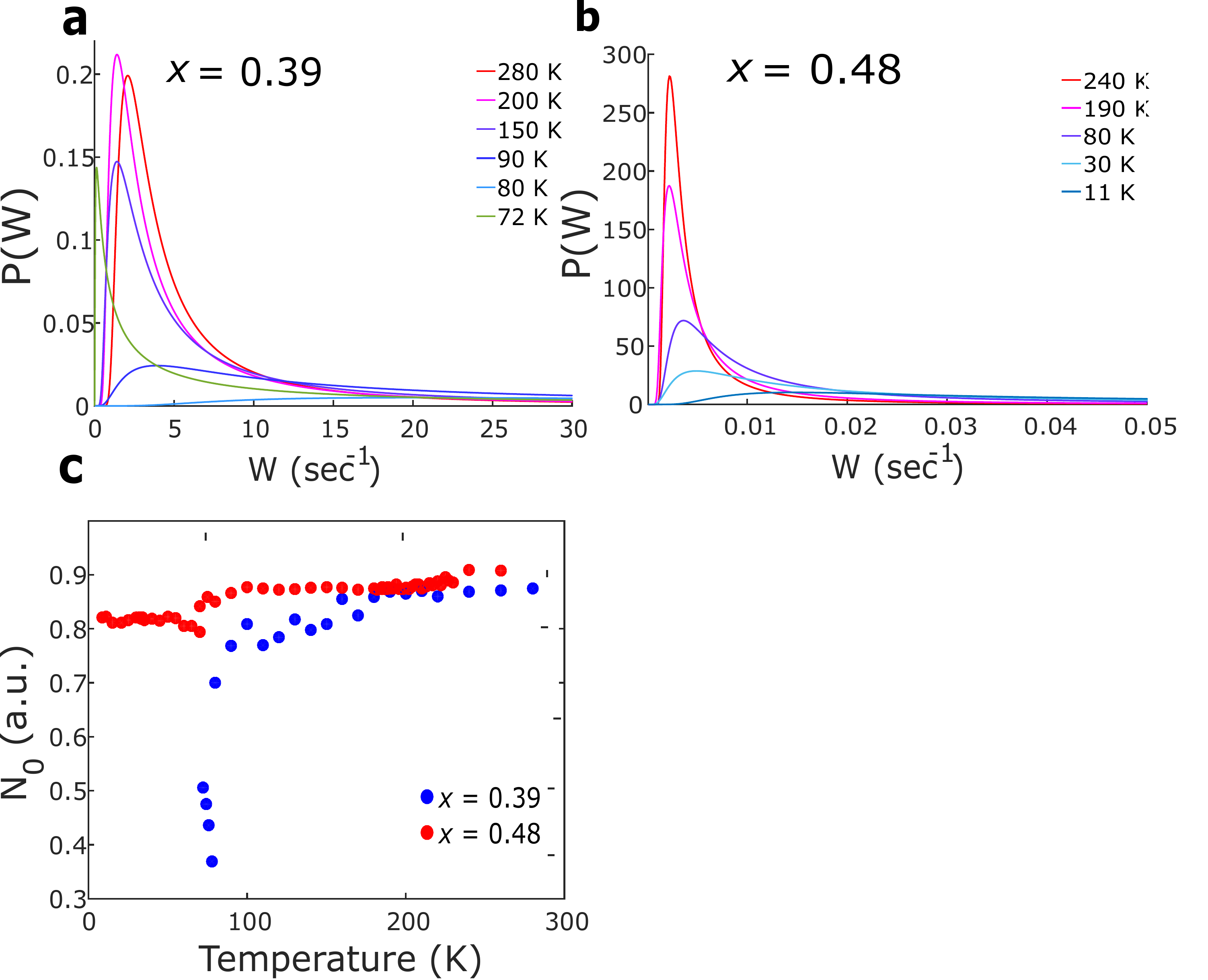}
	\caption{Quantitative study of `wipe-out' effect. (a), (b) Distribution of spin-lattice relaxation rate (1/T$_{1}$), $P(W)$ where $W = 1/T_{1}$, for compounds $x = 0.39$ and 0.48. The distribution of $1/T_{1}$ becomes more asymmetric and develops a long tail to faster rates with decreasing temperature. (c) Temperature dependence of $N_{0}$, the fraction of $^{23}$Na nuclei contributing to the main-Na spectrum. Values of $N_{0}$ are obtained from integrating $P(W)$ for different temperatures up to a cutoff value $W_{cut}$, defined as the rate at which $P(W_{cut})$ drops to 1\% of its maximum~\cite{Mit.08}. For $x = 0.39$, values of $N_{0}$ below 100 K have limited accuracy due to poor signal-to-noise in this temperature range.}\label{wipeout}
\end{figure}
Similar to what has been found in a glassy system such as underdoped cuprates and ``122'' pnictide~\cite{Jul.01,Cur.00,Sin.02,Dio.13,Mit.08}, the total integrated spectral intensity of main-Na peak diminishes dramatically with decreasing $T$ for $x \approx$ $0.39$. This phenomenon, known as `wipe-out' effect, indicates that not all of the nuclei are contributing to the NMR signal due to inhomogeneous and glassy freezing of spin dynamics within the antiferromagnetic (AFM) Fe-Cu clusters\cite{Dio.13,Cur.00}. 

To qualitatively study the `wipe-out' effect, we first derive the $1/T_{1}$ distribution function, $P(W)$ where $W = \frac{1}{T_1}$, for different temperatures with the equations~\cite{Ber.05,Dio.13}:
\begin{equation}
\begin{split}
P(W) = T_{1}\frac{B}{(WT_{1})^{(1-\beta/2)/(1-\beta)}}\textrm{exp}[-\frac{(1-\beta)\beta^{\beta/(1-\beta)}}{(WT_{1})^{\beta/(1-\beta)}}]\\\times\frac{1}{1+C(WT_1)^{\beta(0.5-\beta)/(1-\beta)}} \quad\textrm{(for $\beta$ $\leq$ 0.5)}
\end{split}
\end{equation}
\begin{equation}
\begin{split}
P(W) = T_{1}\frac{B}{(WT_{1})^{(1-\beta/2)/(1-\beta)}}\textrm{exp}[-\frac{(1-\beta)\beta^{\beta/(1-\beta)}}{(WT_{1})^{\beta/(1-\beta)}}]\\\times(1+C(WT_{1})^\frac{\beta(\beta-0.5)}{1-\beta}) \quad\textrm{(for $\beta$ $>$ 0.5)}
\end{split}
\end{equation}
\noindent where values of $B$ and $C$ are $\beta$-dependent and given in Ref.~\cite{Ber.05}. Then we determine, $N_{0}$, the fraction of $^{23}$Na nuclei contributing to the main-Na spectrum, by first integrating $P(W)$ for different temperatures up to a cutoff value $W_{cut}$, defined as the rate at which $P(W_{cut})$ drops to 1\% of its maximum~\cite{Mit.08}. As shown in Fig.~\ref{wipeout}\,(a) and (b), $P(W)$ becomes significantly more asymmetric and develops a long tail extending to faster rates with decreasing temperature, signifying development of inhomogeneity in the system with $T$ reaching $T_{g}$ and below. The temperature dependence of $N_{0}$ plotted in Fig.~\ref{wipeout}\,(c), clearly shows that $N_{0}$ diminishes significantly with decreasing temperature for $x = 0.39$, consistent with the $^{23}$Na spectra. 

\newcommand{\beginappendixb}{%
	\setcounter{table}{0}
	\renewcommand{\thetable}{B\arabic{table}}%
	\setcounter{figure}{0}
	\renewcommand{\thefigure}{B\arabic{figure}}%
	\setcounter{equation}{0}
	\renewcommand{\theequation}{B\arabic{equation}}%
}
\beginappendixb
\section{APPENDIX B: Room-temperature $^{75}$As Spectra for \textit{\textbf{H}}$_{0}$ $||$ $c$-axis}
\begin{figure}[h]
	\hspace*{-0.5cm}
	\includegraphics[scale = 0.33]{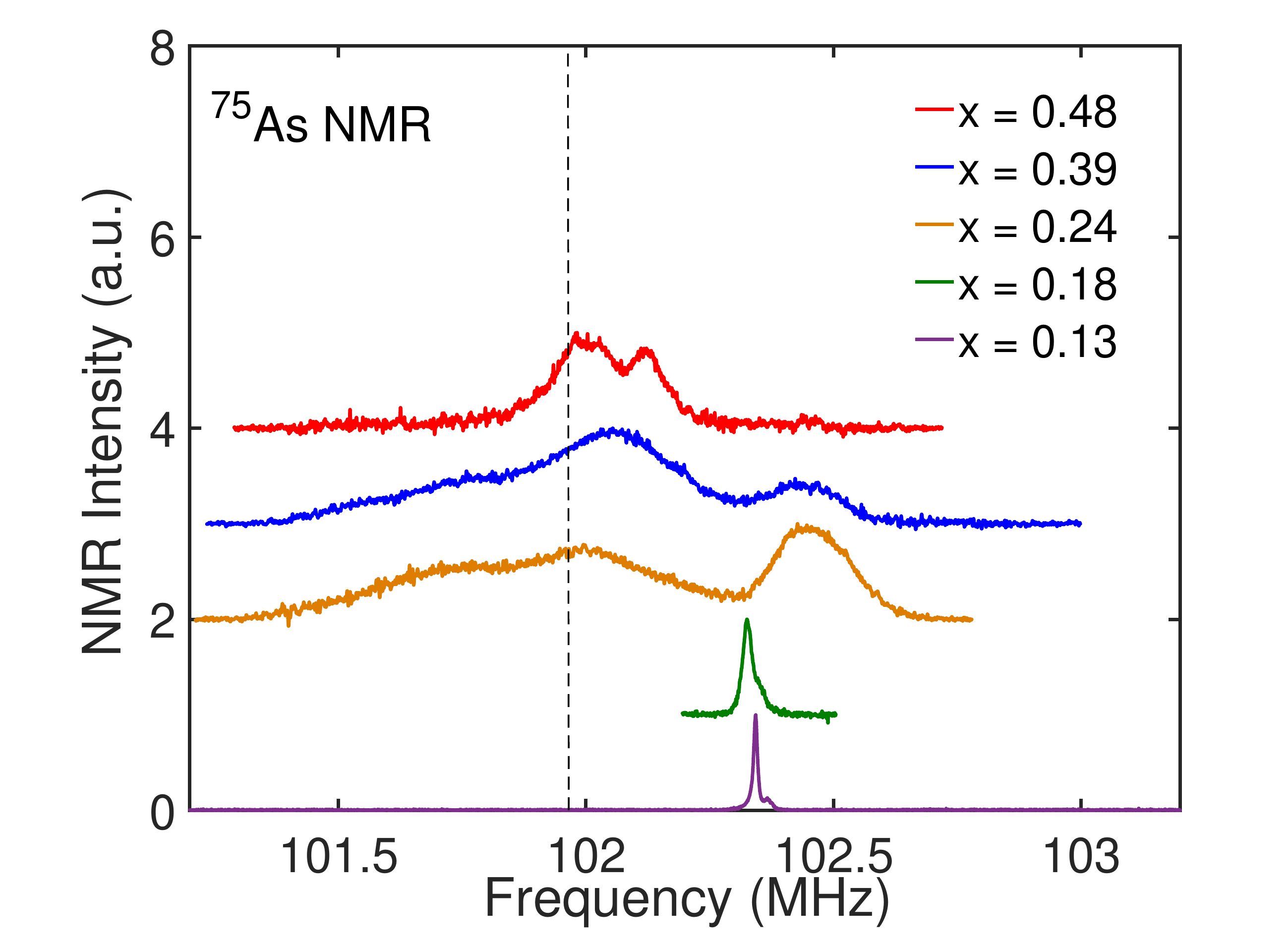}
	\caption{Doping evolution of room-temperature $^{75}$As NMR spectra (\textbf{\textit{H}}$_{0}$ = 14 T $\parallel$ $c$-axis). The dashed line is the Larmor frequency. All spectra are normalized to peak height.}\label{spectra}
\end{figure}

The room-temperature $^{75}$As spectra are shown in Fig.~\ref{spectra}, for $x$ = 0.13, 0.18, 0.24, and 0.39, with \textit{\textbf{H}}$_{0}$ = 14 T $||$ $c$-axis. The dashed line is the Larmor frequency. For $x\leq 0.18$, the $^{75}$As spectra resembles that of $^{75}$Na and show inequivalent $^{75}$As sites having different number of nearest-neighbor (NN) occupied by a Cu dopant. For compounds with Cu concentration $x \geq 0.24$, however, a spectral component centered close to the Larmor frequency starts to develop with increasing $x$ and can be attributed to As sites that are within the Fe-Cu stripes. Since the 2$^{\mathrm{nd}}$ order quadrupolar contribution to the $^{75}$As central transition is negligible, the significant difference between the $^{75}$As spectra for compounds with $x$ close to 0.5 and those with low doping suggests new evolution of local magnetic environments in the paramagnetic state as $x$ is increased beyond 0.24. The most striking feature of the $^{75}$As spectra is the splitting $\sim$130 KHz for $x = 0.48$ above $T_{N}$. Detailed interpretation of the splitting with a staggered magnetization model is discussed in the main text.

\newcommand{\beginappendixc}{%
	\setcounter{table}{0}
	\renewcommand{\thetable}{C\arabic{table}}%
	\setcounter{figure}{0}
	\renewcommand{\thefigure}{C\arabic{figure}}%
	\setcounter{equation}{0}
	\renewcommand{\theequation}{C\arabic{equation}}%
}
\beginappendixc
\section{APPENDIX C: Simulation of Fe-Cu Lattice with Stripe Order}

\begin{figure}[H]
	\includegraphics[scale = 0.32]{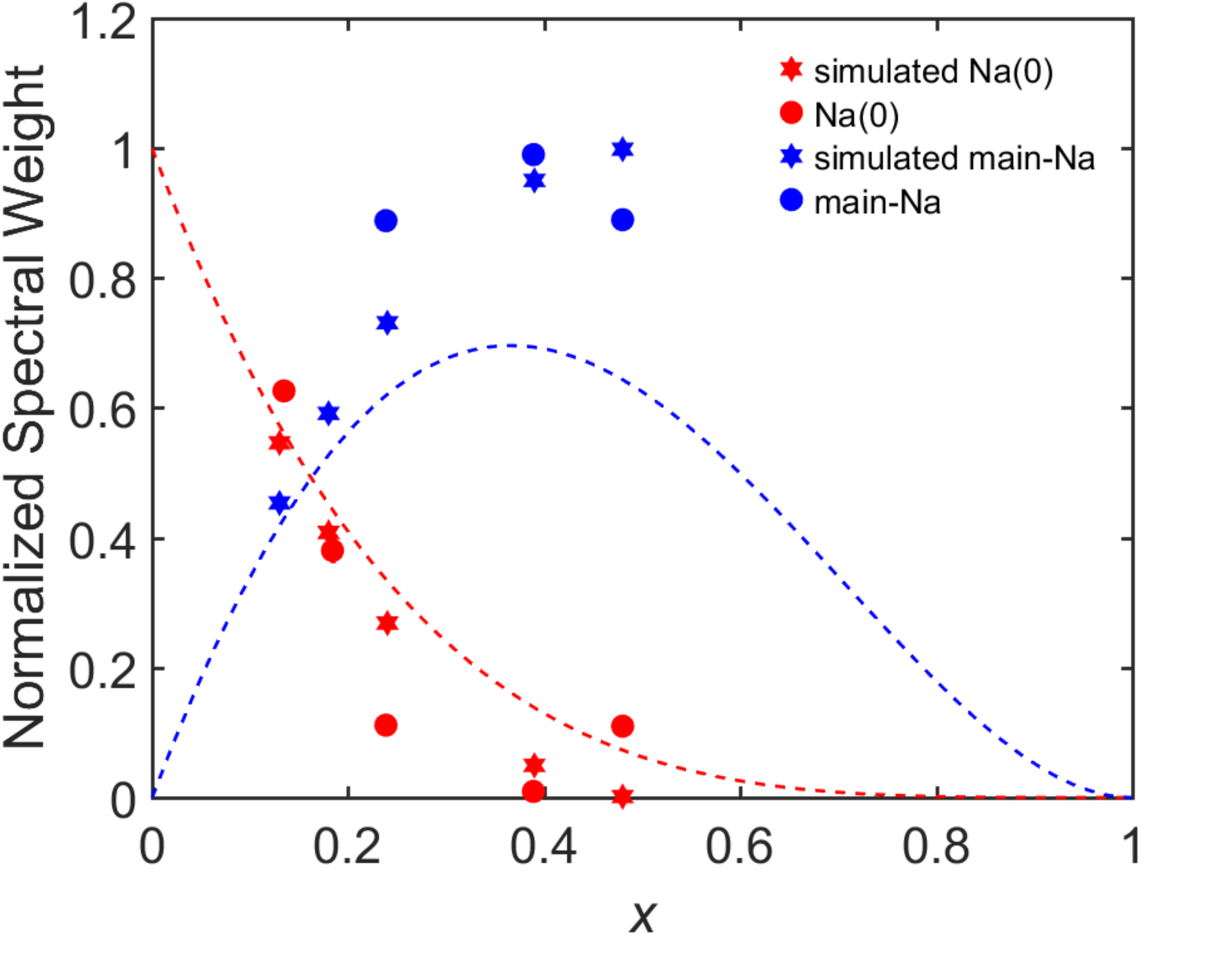}
	\caption{Comparison of relative spectral weights of $^{23}$Na room-temperature spectra and those of Na(0) and main-Na sites from simulation. The dashed lines represent spectral weight, given by a binomial model intended for a random distribution of Cu; these deviate from the measurement results for $x\geq0.24$, while our simulated lattice gives rise to spectral weights that agree reasonably well with the experimental values, suggesting that our scheme of creating the simulated lattice captures the stripe-forming feature in the compounds with $x$ close  to 0.5.}\label{latticesimulation}
\end{figure}

As shown in Fig.~\ref{latticesimulation}, for $x\leq 0.18$, the relative spectral weight of both Na(0) and main-Na sites agree well with that given by a binomial model (dashed line), indicating a random distribution of Fe and Cu~\cite{Xin.19}. With increasing $x$, however, Fe and Cu start to form stripes instead of being randomly populated, evident from the deviation of the spectral weight of main-Na from that given by the binomial model. We show that the spectral weight of main-Na and Na(0) sites both agree reasonably well with that given by our 400 $\times$ 400 simulated lattice for $x$ close to 0.5. 

\newcommand{\beginappendixd}{%
	\setcounter{table}{0}
	\renewcommand{\thetable}{D\arabic{table}}%
	\setcounter{figure}{0}
	\renewcommand{\thefigure}{D\arabic{figure}}%
	\setcounter{equation}{0}
	\renewcommand{\theequation}{D\arabic{equation}}%
}
\beginappendixd
\section{APPENDIX D: Numerical Simulation of $^{75}$As Spectra}

\begin{figure}[h]
	\includegraphics[scale = 0.80]{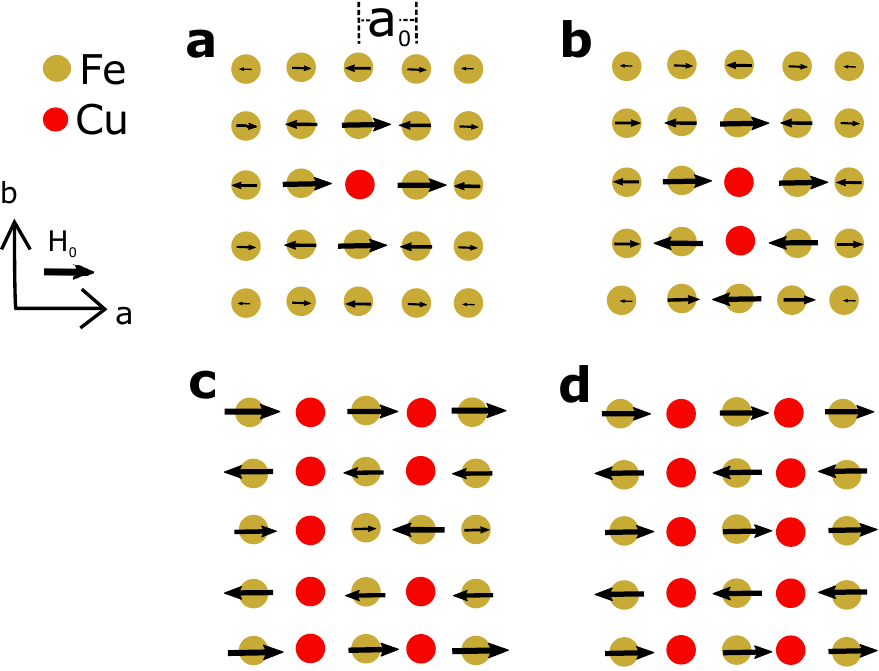}
	\hspace*{+0.5cm}

\caption{Schematic of staggered magnetization peaked at \textbf{\textit{Q}}$_\mathrm{AF}$ induced around non-magnetic Cu-dopants in the presence of an external magnetic field \textbf{\textit{H}}$_{0}$. The progression for increasing Cu concentration toward x = 0.5 is shown for a) through d).
The   direction and magnitude of the staggered magnetization and the magnetic field \textbf{\textit{H}}$_{0}$ is schematic. In actuality, they point along the  \textit{\textbf{c}}-axis.}\label{simulationphysics}
\end{figure}

\begin{figure}[h]
	\includegraphics[scale = 0.50]{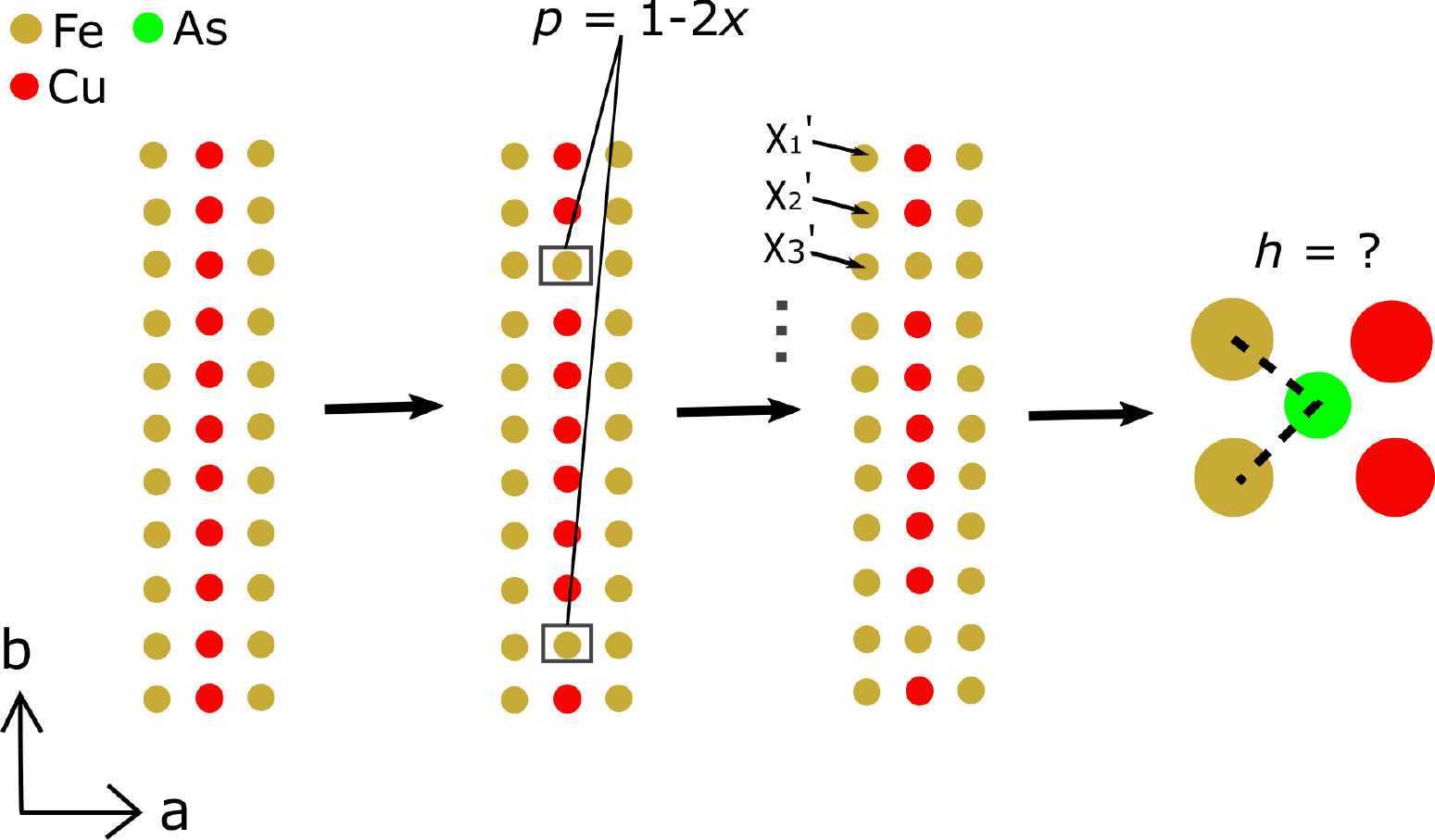}
	\caption{Step by step procedure for numerical simulation of $^{75}$As spectra. We first simulate the Fe-Cu lattice by populating a 400$\times$400 lattice with alternating chains of Fe and Cu. Then we introduce a Fe atom (defect) with a probability $p = 1 - 2x$ at each Cu site. We calculate the total susceptibility $\chi^{'}_{i}$ from all Cu atoms at each Fe site positioned at \textbf{\textit{r$_i$}} in the lattice. Finally we calculate the hyperfine field, $h$, at each As site transferred from nearest-neighbor Fe sites to obtain the As spectrum.}\label{simulationprocedure}
\end{figure}

The non-magnetic dopant-induced enhancement of AFM correlation peaked at \textbf{\textit{Q}}$_\mathrm{AF}$ lies at the foundation of our numerical simulation of $^{75}$As spectra. A staggered magnetization \textbf{\textit{M}}$_{i}$, in the presence of an external magnetic field \textbf{\textit{H}}$_{0}$, at the Fe site position \textbf{\textit{r}}$_{i}$, can be obtained from \textbf{\textit{M}}$_{i}$ = \textbf{\textit{H}}$_{0}$$\chi^{'}_{i}$, where $\chi^{'}_{i}$ is the staggered susceptibility given by Eq.~\ref{Eq.3} summed over all Cu atoms. A progression with increased doping toward x = 0.5 is depicted in Fig.~\ref{simulationphysics}.
	
In a lattice occupied by ideal Fe-Cu stripes without defects, as shown in Fig.~\ref{simulationphysics}\,(d), the hyperfine field along the $c$-axis at As sites is negligible by symmetry. However, as shown in Fig.~\ref{simulationphysics}\,(c), a Fe defect introduces an uncompensated hyperfine field at As sites since the hyperfine field transferred from their nearest-neighbor Fe atoms no longer cancel each other out along the $c$-axis.  This results in an alternating hyperfine field at  As sites near these end-chain defects (as shown in Fig.~\ref{Fig.4}\,(b)), and causes the splitting of the spectrum. A step-by-step guide to the simulation is outlined in Fig.~\ref{simulationprocedure}.

A connection can be made with NMR studies of staggered magnetization induced by both non-magnetic and magnetic impurities in cuprates~\cite{Jul.00,Bob.97,Wal.93, Mor.98}. The enhancement of AFM correlations by non-magnetic dopants is a common phenomenon in both pnictides and cuprates.

\newcommand{\beginappendixe}{%
	\setcounter{table}{0}
	\renewcommand{\thetable}{E\arabic{table}}%
	\setcounter{figure}{0}
	\renewcommand{\thefigure}{E\arabic{figure}}%
	\setcounter{equation}{0}
	\renewcommand{\theequation}{E\arabic{equation}}%
}
\beginappendixe
\section{APPENDIX E: $^{75}$As Spectrum Simulation for $x = 0.48$ with \textbf{\textit{Q}}$_\mathrm{AF}$ = \textrm{(0,1,$\frac{1}{2}$)}}
\begin{figure}[h]
	\includegraphics[scale = 0.35]{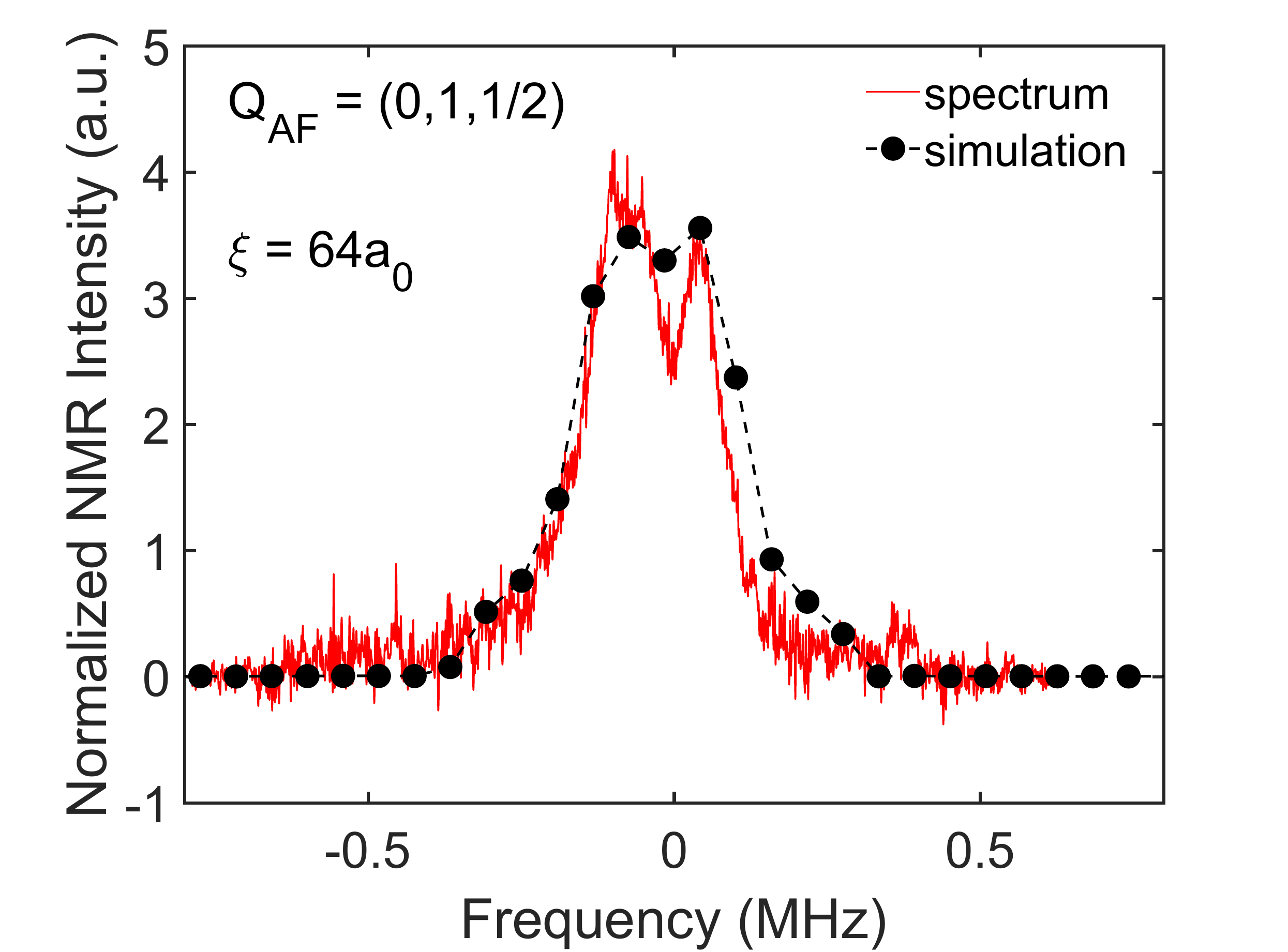}
	\caption{$^{75}$As spectrum simulation on a 400$\times$400 lattice for $x = 0.48$ with \textbf{\textit{Q}}$_\mathrm{AF}$ = (0,1,$\frac{1}{2}$). The simulation results are similar to that with \textbf{\textit{Q}}$_\mathrm{AF}$ = (1,1,0): the correlation lengths $\xi$ agree with one another within uncertainty.}\label{spectrasimulation2}
\end{figure}

Additional simulations are done for \textbf{\textit{Q}}$_\mathrm{AF}$ = (0,1,$\frac{1}{2}$), for the compound $x = 0.48$. For \textbf{\textit{Q}}$_\mathrm{AF}$ = (0,1,$\frac{1}{2}$), the susceptibility at a Fe site at $\textbf{\textit{r}}_i$ with respect to a Cu dopant, $\chi^{'}(\textbf{r}_i-\textbf{r}_\mathrm{Cu})$, becomes

\begin{equation}\label{Eq.4}
\chi^{'}(\textbf{r}_i-\textbf{r}_\mathrm{Cu}) = \chi^{*}(-1)^{n_b}\mathrm{exp}(-|\textbf{r}_i-\textbf{r}_\mathrm{Cu}|^2/4\xi^2)
\end{equation}
Changing from $(-1)^{n_{a}+n_{b}}$ to $(-1)^{n_{b}}$ flips the direction but maintains the magnitude of the spin polarization revealed by a Fe defect in the Cu chain at the Fe sites in the Fe chains. This change has the same effect on the hyperfine field at the majority of the As sites; therefore, the simulated $^{75}$As spectrum and fitting results for \textbf{\textit{Q}}$_\mathrm{AF}$ = (0,1,$\frac{1}{2}$) are similar to that for  \textbf{\textit{Q}}$_\mathrm{AF}$ = (1,1,0), as shown in Fig.~\ref{spectrasimulation2}.


\bibliography{maintext}

\end{document}